\newcommand{\BaCdV} {BaCdVO(PO$_4$)$_2$\xspace}

\newcommand{\be}{\begin{equation} }
\newcommand{\ee}{\end{equation} }
\newcommand{\bea}{\begin{eqnarray} }
\newcommand{\eea}{\end{eqnarray} }

\documentclass[aps,prb,reprint,showpacs,longbibliography,superscriptaddress, citeonline]{revtex4-2}
\usepackage{amsmath,amssymb,bm}
\usepackage{graphicx}
\usepackage{xspace}
\usepackage{latexsym}
\usepackage{color}
\usepackage{hyperref}
\usepackage{placeins}

\newcommand{\BaCd}{BaCdVO(PO$_4$)$_2$\xspace}

\newcommand{\PbV}{Pb$_2$VO(PO$_4$)$_2$\xspace}

\begin{document}

\title{Inelastic neutron scattering determination of the spin Hamiltonian for \BaCdV}
\author{V.~K.~Bhartiya}
\email{vvivek@phys.ethz.ch}
\affiliation{Laboratory for Solid State Physics, ETH Z\"{u}rich, 8093 Z\"{u}rich, Switzerland}

\author{S.~Hayashida}
\affiliation{Laboratory for Solid State Physics, ETH Z\"{u}rich, 8093 Z\"{u}rich, Switzerland}

\author{K.~Yu.~Povarov}
\affiliation{Laboratory for Solid State Physics, ETH Z\"{u}rich, 8093 Z\"{u}rich, Switzerland}

\author{Z.~Yan}
\affiliation{Laboratory for Solid State Physics, ETH Z\"{u}rich, 8093 Z\"{u}rich, Switzerland}

\author{Y.~Qiu}
\affiliation{NIST Center for Neutron Research, National Institute of Standards and Technology, Gaithersburg, Maryland 20899, USA}

\author{S.~Raymond}
\affiliation{Univ. Grenoble Alpes, CEA, IRIG/MEM-MDN, F-38000 Grenoble, France}
\author{A.~Zheludev}
\email{zhelud@ethz.ch}
\homepage{http://www.neutron.ethz.ch/}
\affiliation{Laboratory for Solid State Physics, ETH Z\"{u}rich, 8093 Z\"{u}rich, Switzerland}

\begin{abstract}
  Single crystal inelastic neutron scattering is used to study spin wave excitations in the fully polarized state of the frustrated quantum ferro-antiferromagnet \BaCd. The data analysis is based on a Heisenberg spin Hamiltonian that includes as many distinct nearest-neighbor and next-nearest neighbor interactions as allowed by crystal symmetry. All 8 such exchange constants are obtained in a  simultaneous fit to over 150 scans across  the dispersion manifold. This establishes a definitive quantitative model of this material. It turns out to be substantially different from the one assumed in numerous previous studies based on powder experiments.
\end{abstract}

\date{\today}
\maketitle
\section{Introduction}

Despite its apparent simplicity, the square lattice $S=1/2$ Heisenberg model with ferromagnetic (FM) nearest-neighbor (NN) coupling $J_1$ and frustrating antiferromagnetic (AF) next nearest neighbor (NNN) interaction $J_2$ is among the most important models in magnetism. It is famous for supporting an exotic spin-nematic phase for sufficiently strong frustration ratios \cite{Shannon2006,Shindou2009,Shindou2009,Smerald2015} or  in applied magnetic fields for moderate frustration \cite{Ueda2013,Ueda2015}. Unfortunately, no perfect experimental realizations of this model have been discovered to date. The closest approximations are found among layered vanadyl phosphates with the general formula  AA$^\prime$VO(PO$_4$)$_2$  (A, A$^\prime$ = Ba, Cd, Pb, Sr, Zn) \cite{Nath2008,TsirlinSchmidt2009,TsirlinRosner2009}. Of these the most frustrated and the most promising spin nematic candidate is \BaCd \cite{TsirlinSchmidt2009,TsirlinRosner2009}. Indeed, recent studies have produced compelling thermodynamic and neutron diffraction evidence that this material may have a novel exotic quantum phase in a wide range of applied magnetic fields below saturation~\cite{Povarov2019,Bhartiya2019}.

All initial estimates of the coupling constants and frustration strengths in  AA$^\prime$VO(PO$_4$)$_2$ materials were based on powder sample experiments analyzed with the assumption that the underlying model is indeed a perfect $J_1$-$J_2$ square lattice \cite{Nath2008}. However, the latter is incompatible with the crystal symmetries of any compound in the family. All evidence points to that NN interactions stay FM, NNN remain AFM, but beyond that the deviations from simple square lattice symmetry are substantial.  For example, single crystal experiments on  \PbV revealed that it has as many as 5 distinct exchange constants and a weaker frustration than suggested by previous powder studies \cite{Bettler2019,Landolt2020}. The situation is even more complicated for \BaCd,  where the powder/perfect square lattice estimate is $J_2/J_1=-0.9$~\cite{Nath2008}. Already the room temperature crystal structure \cite{Meyer1997} allow for 4 distinct exchange constants. A recently discovered structural transition \cite{Bhartiya2019} at 240~K lowers the symmetry even further. As many as 4 nearest-neighbor and 4 next nearest neighbor coupling constants are allowed.

The main question is whether the rather complex interactions in \BaCd are compatible with the presence of a high-field nematic state. To answer it one has to know the exact values of exchange parameters. A first step in this direction was made in our preliminary inelastic neutron scattering study of the spin wave spectrum in the fully saturated phase \cite{Bhartiya2019}. Due to the limited amount of data that could be collected on a unique but small $^{114}$Cd enriched crystalline sample,  it was not possible to determine all 8 relevant parameters unambiguously. Nonetheless, the data were enough to demystify the peculiar ``up-up-down-down" zero field magnetic structure previously detected in powder experiment \cite{Skoulatos2019}. As for the field-induced nematic phase, recent theoretical calculations made use of our preliminary estimates to demonstrate its robustness \cite{Smerald2020}. Still, the exact Hamiltonian remains undetermined.

In the present work we report the results of a full-scale continuation of the preliminary neutron measurement. We utilize the extremely high efficiency MACS spectrometer at the National Institute of Standards (NIST) to map out the spin wave dispersion in the entire Brillouin zone. We then analyze the combination of new and previously collected data in a single global model fit. In doing so we fully take into account the complex mosaicity of the sample and the energy-momentum resolution of the spectrometers. The result is a definitive spin Hamiltonian for \BaCd.

  \begin{figure*}
  \includegraphics[width=1\textwidth]{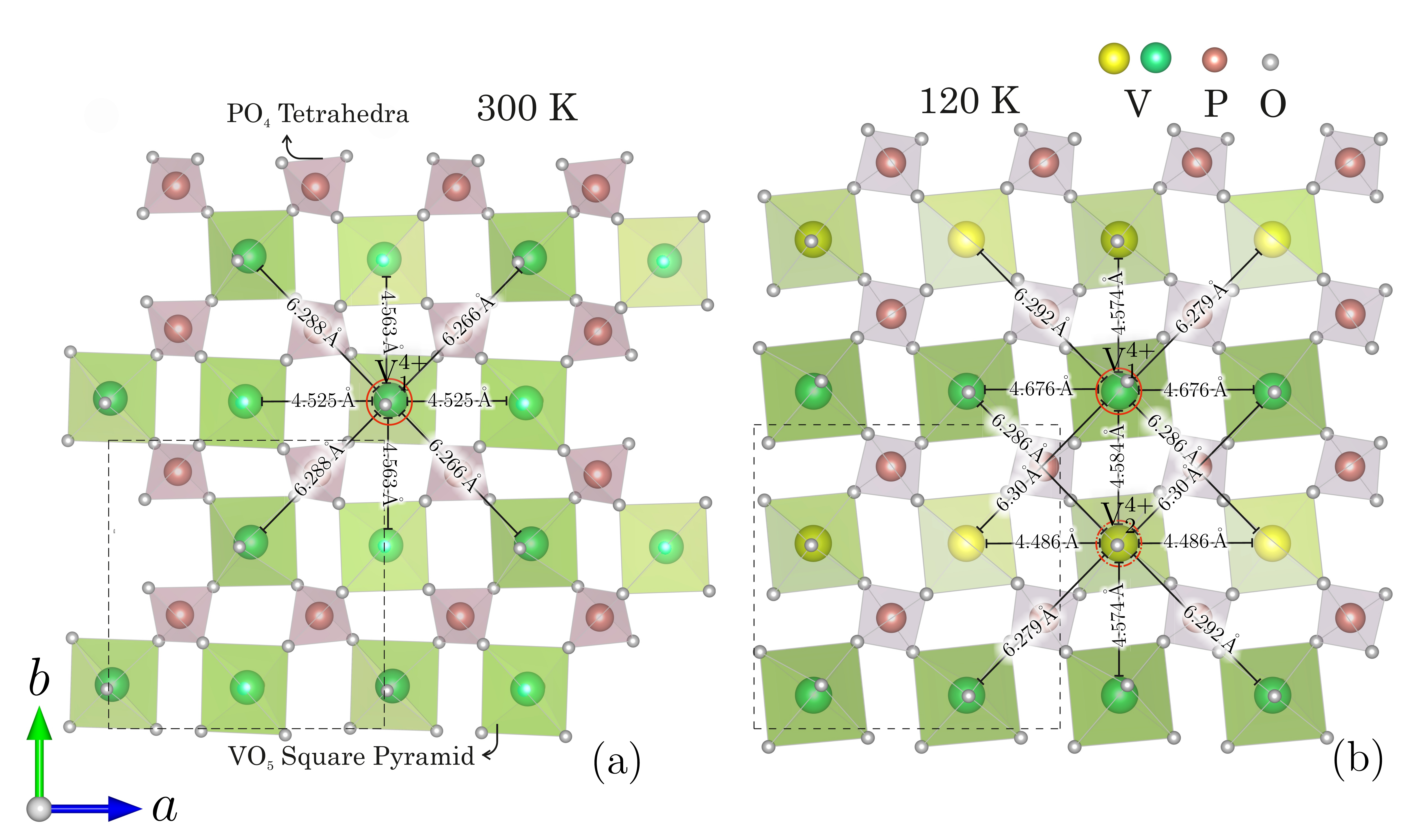}\\
  \caption{Structure of vanadyl phosphate layers in \BaCd at 300~K in the $Pbca$ phase (a) and at 120~K in the $Pca2_1$ phase (b). Distances between nearest- and next-nearest magnetic V$^{4+}$ ions are indicated.  The two inequivalent V$^{4+}$ are shown with different color. The dotted black rectangle shows the crystallographic unit cell. }\label{FIG:xtalPT}
  \end{figure*}
 

\section{Material and Experimental Details}
The room temperature crystal structure of \BaCd is orthorhombic (space group $Pbca$ No. 61) with lattice parameters $a=8.84$ \AA, $b=8.92$ \AA, and $c=19.37$ \AA ~\cite{Meyer1997}. Magnetism is due to $S=1/2$ V$^{4+}$ ions that form layers parallel to the $(a,b)$ plane. There are 8 magnetic ions per crystallographic unit cell, four  in each layer within a single cell. Intra-layer NN and NNN interactions are expected to dominate.  Further in-plane and inter-layer coupling are expected to be negligible \cite{TsirlinRosner2009}. In particular, the spin wave dispersion measured previously in the very similar \PbV system is perfectly modeled without taking these into account \cite{Bettler2019}. Already at room temperature there are two distinct NN and two NNN superexchange pathways, as illustrated in Fig.~\ref{FIG:xtalPT}a, which shows a single magnetic layer.  As mentioned above, the crystal symmetry is further lowered upon cooling through a structural transition at about 250~K. At $T=120 $~K the space group is $Pca2_1$ ($C_{2\nu}^{5}$, No. 29) with lattice parameters $a = 8.8621(4)$ \AA, $b = 8.8911(4)$ \AA, and $c = 18.8581(9)$ \AA  ~\cite{Bhartiya2019}. There are two V$^{4+}$ symmetry-inequivalent sites represented now by two different colors in each layer and 8 distinct superexchange paths as shown in Fig.~\ref{FIG:xtalPT}(b). Magnetic order sets in at  $T_N \simeq 1.05$~K \cite{Povarov2019}.  Its ``up-up-down-down" character \cite{Skoulatos2019} is enabled by the alternation of NN interaction strengths along the crystallographic $a$ axis \cite{Povarov2019}.  A spin-flop transition observed at $\sim 0.5$~T for a field applied along the same direction \cite{Povarov2019} suggest a tiny easy-axis magnetic anisotropy of the order of 0.005~meV. As mentioned, the magnetic phase diagram includes an extensive pre-saturated spin-nematic candidate phase, as discussed in detail in Refs.~\cite{Povarov2019,Bhartiya2019}. Full saturation is reached at $\mu_0 H_\mathrm{sat}\simeq 6.5$~T.

The present measurement of the spin Hamiltonian is based on the method pioneered by Coldea {\em et al.} in Ref.~\cite{Coldea2002}. Inelastic neutron scattering is used to measure the spin wave dispersion in the fully saturated phase. It is then analyzed in the framework of spin wave theory, which for the Heisenberg model becomes exact above saturation. We made use of the same $\sim$320~mg 98\% $^{114}$Cd-enriched sample as in the experiments reported in Ref.~\cite{Bhartiya2019}. In habit it is green and transparent. The synthesis procedure is as follows. Single-phase polycrstalline \BaCd was prepared by the solid-state reaction method in two steps. First, stoichiometric amounts of precursors NH$_4$H$_2$PO$_4$, BaCO$_3$ and $^{114}$CdO were sintered in a Pt crucible at 700$^\circ$~C for 72 hrs to yield single-phase BaCdP$_2$O$_7$. In a second step, the product was  mixed with stoichiometric amounts of V$_2$O$_5$ and V$_2$O$_3$, then compacted under hydrostatic pressure of 70~MPa for 20 minutes. The resulting pellets were sintered  at 800$^\circ$~C for 48 hrs in a glassy graphite crucible sealed in quartz under a vacuum of 10$^{-4}$~Torr. In all cases 99.99$\%$-purity starting materials were used. The crystal was grown from finely ground powders using the self-flux Bridgman technique at  0.2~mm/hr at 880$^\circ$~C in a sealed glassy graphite crucible with tantalum as oxygen scavenger. Powder and single crystal X-ray diffraction experiments in all cases indicate a single phase with no disorder of any kind. The lack of disorder is further supported by a total lack of Curie-like contribution to magnetic susceptibility at low temperatures \cite{Povarov2019}. 

\begin{figure}
\includegraphics[width=0.45\textwidth]{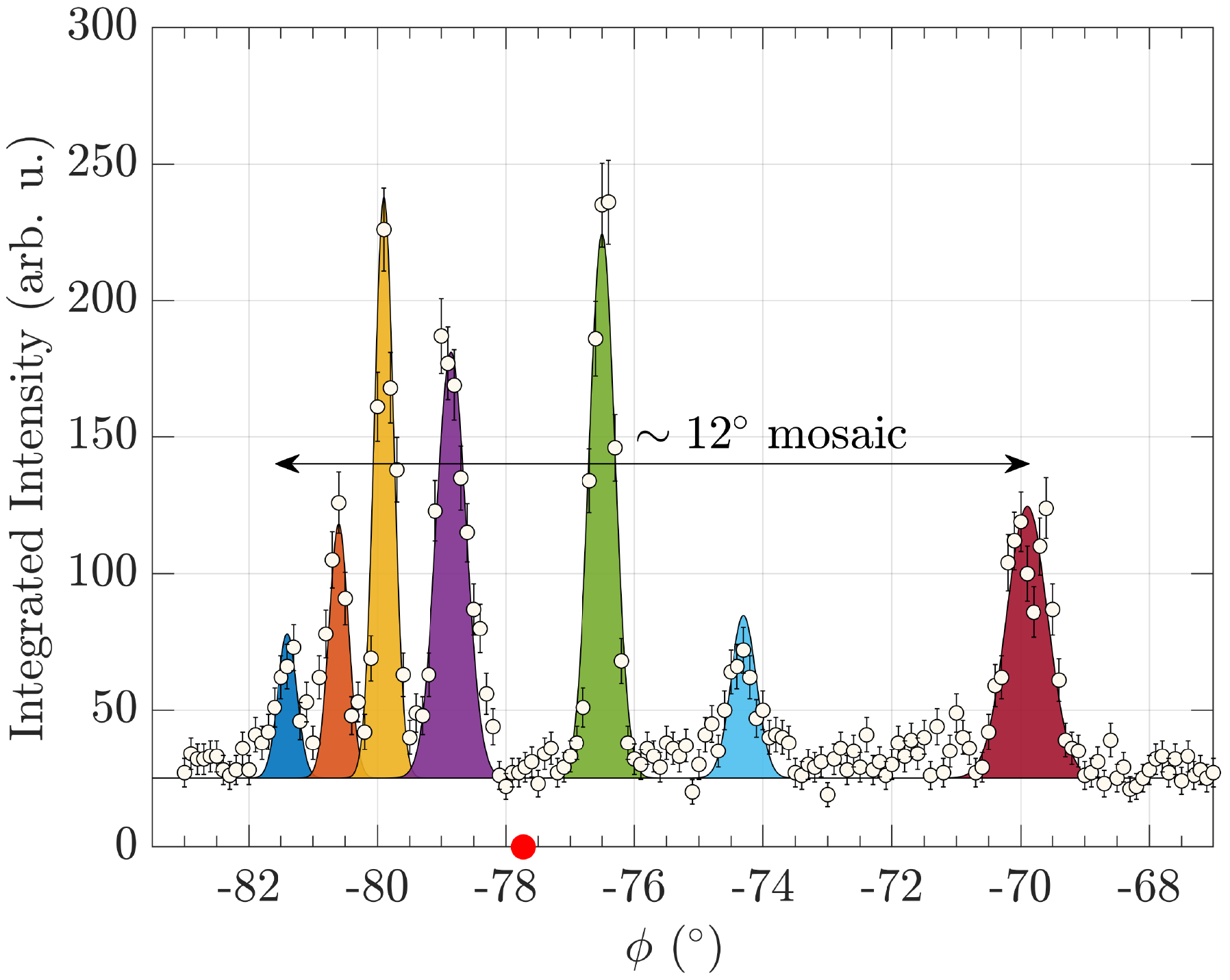}\\
\caption{Tilt-integrated rocking curve (intensity vs. sample rotation angle $\phi$) of the $(0,2,0)$ Bragg peak measured in the \BaCd sample studied in this work and the error bars represent one standard deviation. Contribution of 7 individual crystallites are color-coded. The red dot indicates the position of  center of mass, relative to which all momentum transfers were indexed. }\label{FIG:CrystalAssembly}
\end{figure}

All neutron measurements were carried out with momentum transfers in the $(h,k,0)$ reciprocal space plane. The mosaic of the crystal was characterized by mapping out the distributions of the $(200)$ and $(020)$ Bragg peaks both within and out of the scattering plane using a series of rocking curves. The survey revealed 7 distinct crystallites of individual mosaic spreads $< 1^{\circ}$, but distributed over about $12^{\circ}$ in the $(a,b)$ plane and within $\pm 5^{\circ}$ out of the plane.  A tilt-integrated rocking curve of the $(0 2 0)$ Bragg peak is shown in Fig.~\ref{FIG:CrystalAssembly}.
 An analysis of the measured integrated peak intensities yielded the rotations of individual crystallites in the $(a,b)$ plane relative to the mean  setting ($-6.74^\circ$, $-5.96^\circ$, $-5.19^\circ$, $-4.18^\circ$, $-1.8^\circ$, $0.34^\circ$, $4.8^\circ$), as well as their relative masses (0.13, 0.3, 0.5, 0.65, 0.7, 0.15, 1) normalized to the largest crystallite, correspondingly.

New inelastic data were collected  with the Multi-Axis Crystal Spectrometer (MACS) at National Institute of Standards and Technology (NIST) \cite{Rodriguez2008}. All measurements were done in a 9~T magnetic field applied along the crystallographic $c$ axis. Due to high neutron flux at the neutron absorbing sample the stable temperature was $\sim $700 mK in a $^3$He-$^4$He dilution refrigerator. With it's 20 detectors positioned at different scattering angles but tuned to the same energy, MACS is optimized for measuring two-dimensional intensity maps at a constant energy transfer, as was done, for example, in the study of  \PbV \cite{Bettler2019}. For \BaCd we chose a different approach that allows to better resolve the rather weakly dispersive bands at the bottom of the dispersion manifold \cite{Bhartiya2019}. In our routine one particular detector performed energy scans at fixed wave vectors transfers $\mathbf{q}$ as in a conventional 3-axis experiment:  $(0.1,1.5,0)$, $(0.2,1.5,0)$, $(0.3,1.5,0)$, $(0.4,1.5,0)$, $(0.5,1.5,0)$, $(0.6,1.5,0)$, $(0.7,1.5,0)$, $(0.8,1.5,0)$ and $(1.0,1.5,0)$. The energy was scanned from 0.55~meV to 3~meV with an $E_f= 2.7$~meV fixed-final neutron energy. The energy step was 0.025 meV with counting time of $\sim$ 6 min/point. The measured energy width of the incoherent elastic line was $\sim$ 0.15~meV.

\begin{figure}
\includegraphics[width=0.5\textwidth]{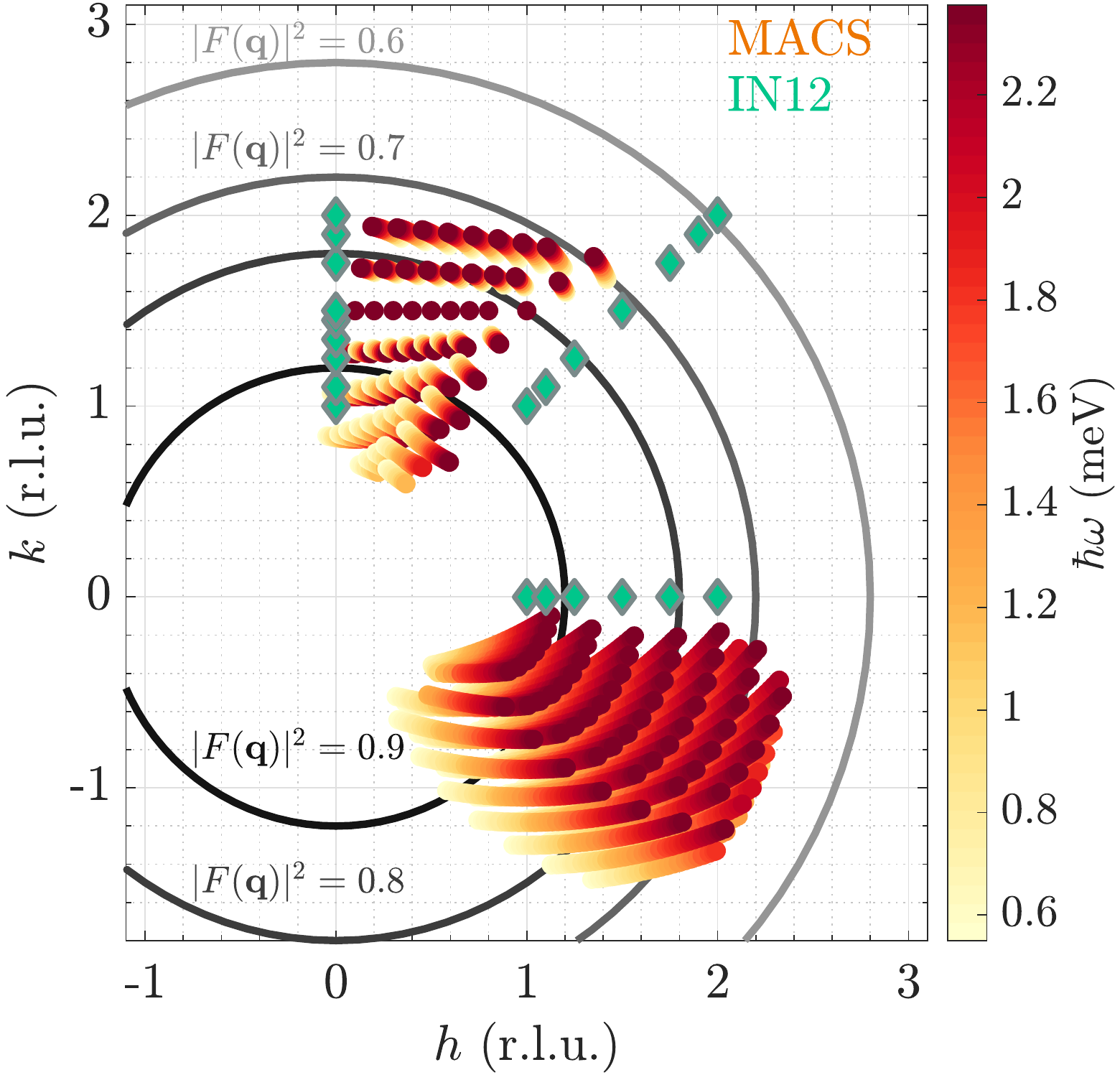}\\
\caption{Coverage of momentum-energy space in inelastic neutron scattering measurements with MACS (circles)  and IN12 ( diamonds, Ref.~\cite{Bhartiya2019}) instruments. The energy is scanned between 0.55 and 3 ~meV. Concentric arcs represent the magnetic form factor squared of V$^{4+}$.}\label{FIG:kspce}
\end{figure}

\begin{figure*}
\includegraphics[width=\textwidth]{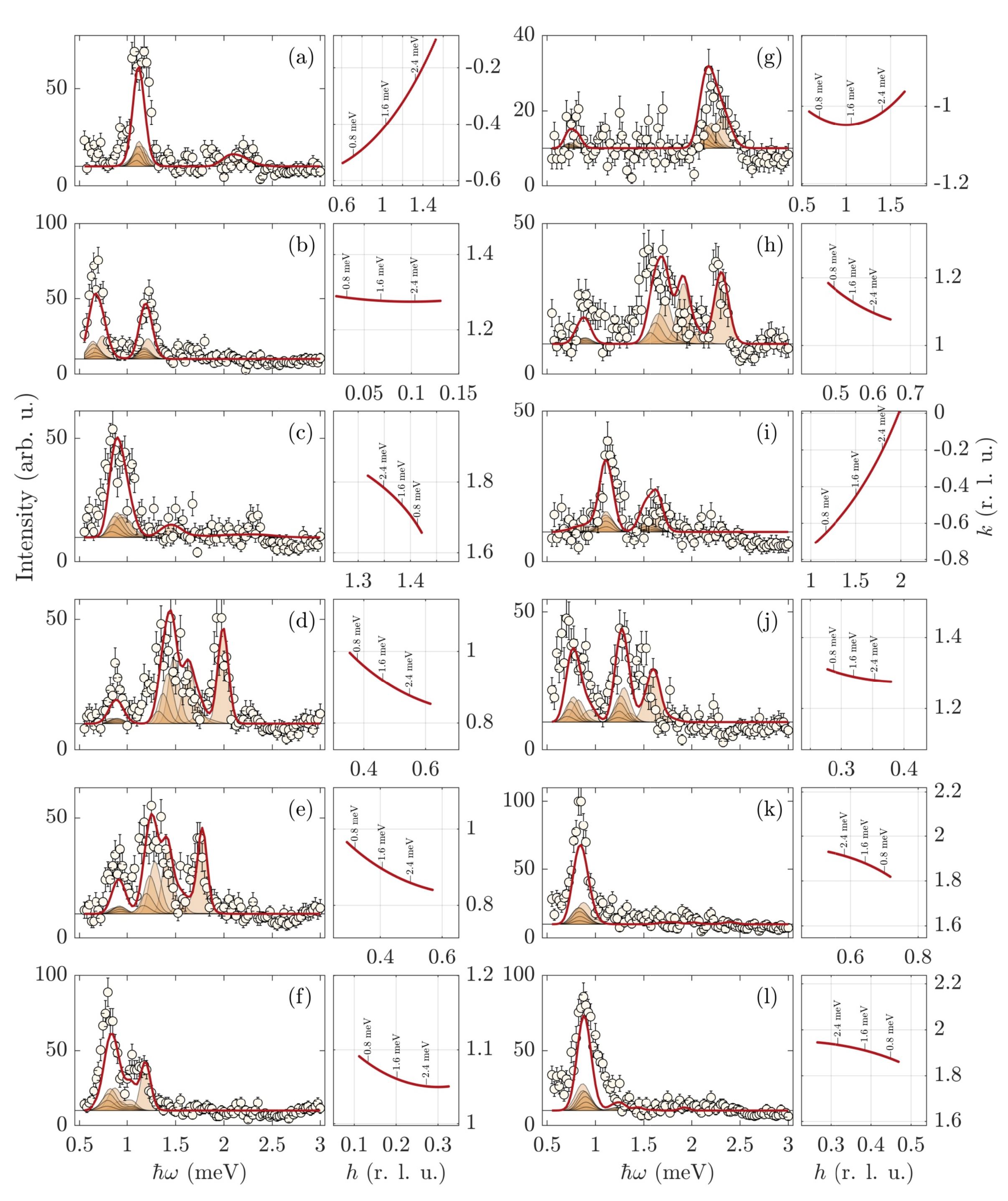}\\
\caption{Left panels: Representative neutron scattering data collected by individual detectors in the course of energy scans on the MACS spectrometer (symbols) and the error bars represent one standard deviation. The solid red line is a result of a global model fit to the entire collected data set, as explained in the text. The shaded peaks are individual contributions of each of the 7 crystallites in the sample.  Right panels: reciprocal space trajectories of the corresponding scans.}\label{FIG:escans}
\end{figure*}

In the course of these constant-$\mathbf{q}$ scans, the remaining 19 detectors  performed ``oblique'' scans in both momentum and energy.  Data points collected with scattering angles below $\pm 20^\circ$ were discarded to avoid background from the direct beam. The final data set consisted of 151 scans of 9510 data points covering a large part of the Brillouin zone, as shown by the colored circles in Fig.~\ref{FIG:kspce}. Representative individual scans are shown in Fig.~\ref{FIG:escans}. In Fig.~\ref{FIG:InelasticScans} we show several representative two-dimensional energy-momentum slices of the collected data. The new MACS data supplemented the data previously collected at the IN12 3-axis spectrometer at ILL in a 10~T  $c$-axis magnetic field \cite{Bhartiya2019}. The latter were all taken in conventional constant-$\mathbf{q}$ scans along high symmetry directions, as indicated by diamond symbols in Fig.~\ref{FIG:kspce}.

\section{Data Analysis}
\begin{figure}
\includegraphics[width=0.45\textwidth]{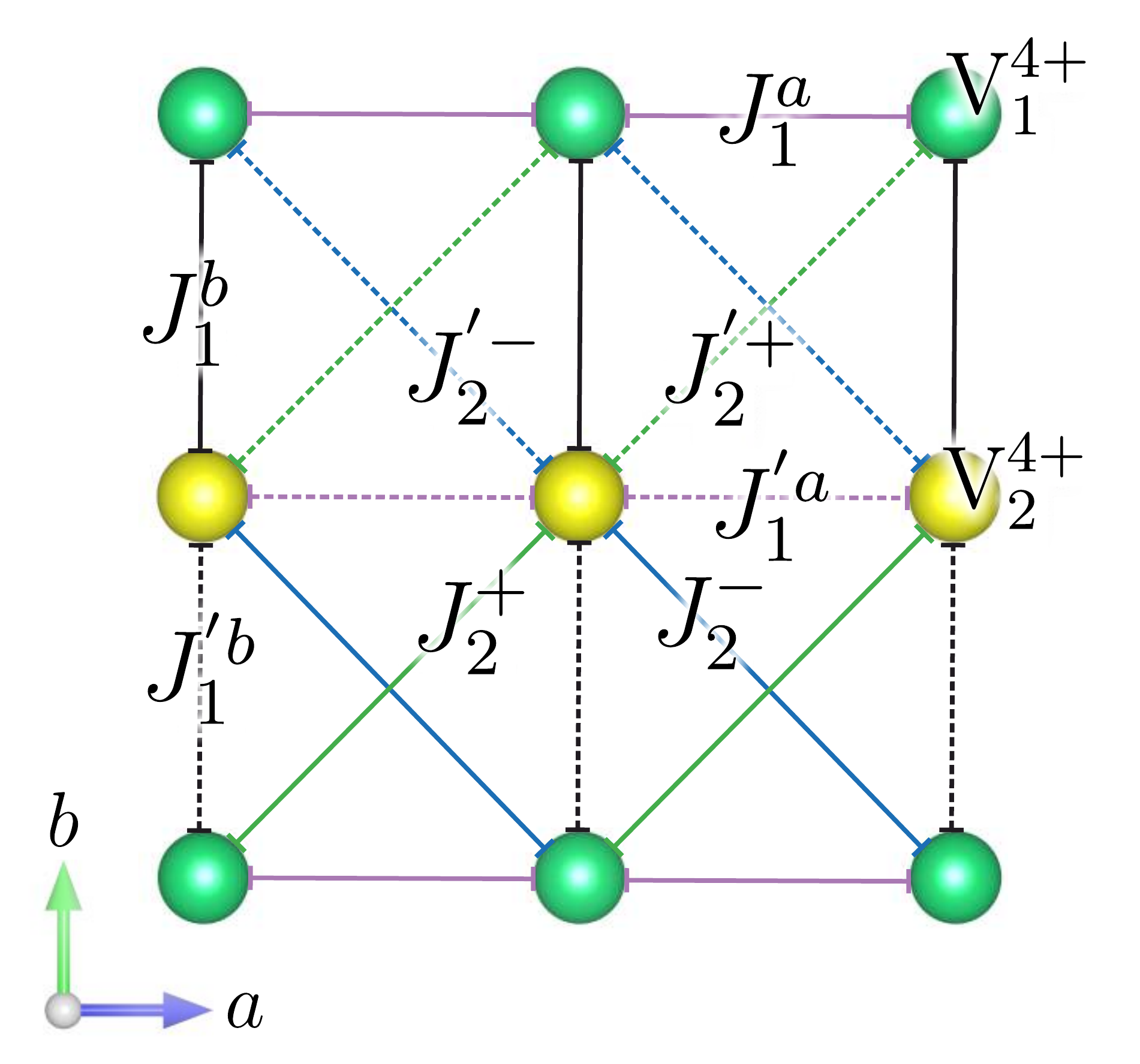}\\
\caption{ Exchange parameters of the Heisenberg model used to analyze the measured magnetic inelastic scattering in \BaCdV.    The two colors of V$^{4+}$ atoms correspond to two symmetry inequivalent sites.} \label{FIG:MagneticStr}
\end{figure}

The analysis of the measured magnetic scattering intensities was based on the Heisenberg model for V$^{4+}$ spin in each layer.  Interactions between layers were assumed negligible. Unlike the constrained model used in Ref.~\cite{Bhartiya2019}, we allowed for 8  distinct exchange constants connecting nearest-neighbor and next-nearest neighbor spins  as shown in Fig. \ref{FIG:MagneticStr}.  To avoid over-parametrization, further in-plane interactions, inter-layer coupling (which is in any case irrelevant for in-plane dispersion in the first order) and anisotropy were assumed negligible, as discussed above.   The 8-parameter spin wave dispersion relation for the fully saturated phase has been worked out in  Ref.~\cite{Smerald2020}. It contains two distinct dispersion branches corresponding to two crystallographically inequivalent V$^{4+}$ sites:
\begin{equation} \label{eq:dispersion}
\hbar\omega_{\mathbf{q}} = \frac{A_{\mathbf{q}} + A'_{\mathbf{q}}}{2} \pm \sqrt{\left(\frac{A_{\mathbf{q}} - A'_{\mathbf{q}}}{2}\right)^2 + |B_{\mathbf{q}}|^2}.
\end{equation}  
Here
\begin{eqnarray}
A'_{\mathbf{q}} & = & \tilde{h}  - J_1^{\prime a}(1- \cos \mathbf{q}\mathbf{a})  \nonumber\\
A_{\mathbf{q}} & = & \tilde{h}  - J_1^{ a}(1- \cos \mathbf{q}\mathbf{a}) \nonumber\\
2 B_{\mathbf{q}} & = & (J_1^{b} e^{i\mathbf{q}\mathbf{b}}+ J_1^{\prime b}  e^{-i\mathbf{q}\mathbf{b}}) +  ( J_2^{+} e^{-i(\mathbf{q}\mathbf{a}-\mathbf{q}\mathbf{b})}\nonumber \\
& +&  J_2^{ \prime +}e^{i(\mathbf{q}\mathbf{a}-\mathbf{q}\mathbf{b})}) + ( J_2^{-} e^{i(\mathbf{q}\mathbf{a}+\mathbf{q}\mathbf{b})} +J_2^{\prime -}e^{-i(\mathbf{q}\mathbf{a}+\mathbf{q}\mathbf{b})}) \nonumber\\
\tilde{h} &=& g\mu_B \mu_0H - \frac{1}{2} (J_1^{b}+J_1^{\prime b} + J_2^{+} +J_2^{ -} + J_2^{\prime +} +J_2^{\prime -}). \nonumber
\end{eqnarray}

 Due to the corrugated character of the V$^{4+}$ layers, each of these branches will give rise to three additional ``replicas'', similarly to  what was seen for zig-zag spin chains   PbNi$_2$V$_2$O$_8$ \cite{Zheludev2000} and  BaCu$_2$Si$_2$O$_7$ \cite{ZheludevKenzelmann2001}. Fortunately, for momentum transfers in the $(h,k,0)$ plane as those explored in the present experiment only the two principal magnon branches are visible. The downside is that any permutations of exchange constants that leave the dispersion relation \eqref{eq:dispersion} intact ($J_1^b\leftrightarrow J_1^{\prime b}$, $J_2^+\leftrightarrow J_2^{\prime +}$,  $J_2^{-}\leftrightarrow J_2^{\prime -}$ and/or $J_1^a\leftrightarrow J_1^{\prime a}$)  cannot be distinguished from the analysis of in-plane data.

The new data collected on MACS was analyzed together with data previously obtained on  IN12 \cite{Bhartiya2019} in a {\em single global fit}. At every wave vector, the energies and intensities of the two magnon branches were numerically computed using the SpinW Matlab library \cite{Toth2015} using the 8 exchange parameters of the model. This was done for each of the 7 crystallites, taking into account their orientations and relative masses, resulting in a total of 14 observable modes.  Neutron intensities were modeled by numerically convoluting this result with the energy-momentum resolution of the spectrometers, the latter calculated in the Popovici approximation~\cite{Popovici1975}. The computation was performed using the ResLib software package ~\cite{Zheludev_ResLib}.  For each of the two experiments we used separate overall intensity prefactors and flat backgrounds. Thus, there are a total of 12 independent parameters in the intensity model.

Even in the $^{114}$Cd-enriched sample the estimated neutron penetration depth is only about 15~mm. This results in intensity attenuations between 15~\% and 35~\% due to absorption depending on scattering geometry and neutron energy.  An exact correction for absorption was unfeasible due to irregular shape of the sample and  unknown spatial distribution of the 7 crystallites. Instead, absorption effects were simply ignored. This approximation is acceptable since the variation of attenuation is estimated to be no more than 10~\% between different data points.

The model was fit to the bulk of experimental data from MACS and IN12 using a Levenberg-Marcquardt least squares procedure. Randomly sampling the initial parameter values consistently produced the same final fit result with good convergence. In the best fit we obtain  $\chi^2=3.05$. Considering the numerous experimental complications and the global nature of the fit, the degree of agreement is very good.   The fitted exchange constants with 95$\%$ confidence interval are listed in Table~\ref{tab:exchange}.   Once again we note that these values are valid only to within the above-mentioned permutations that leave the dispersion intact.

\begin{figure*}
\includegraphics[width=\textwidth]{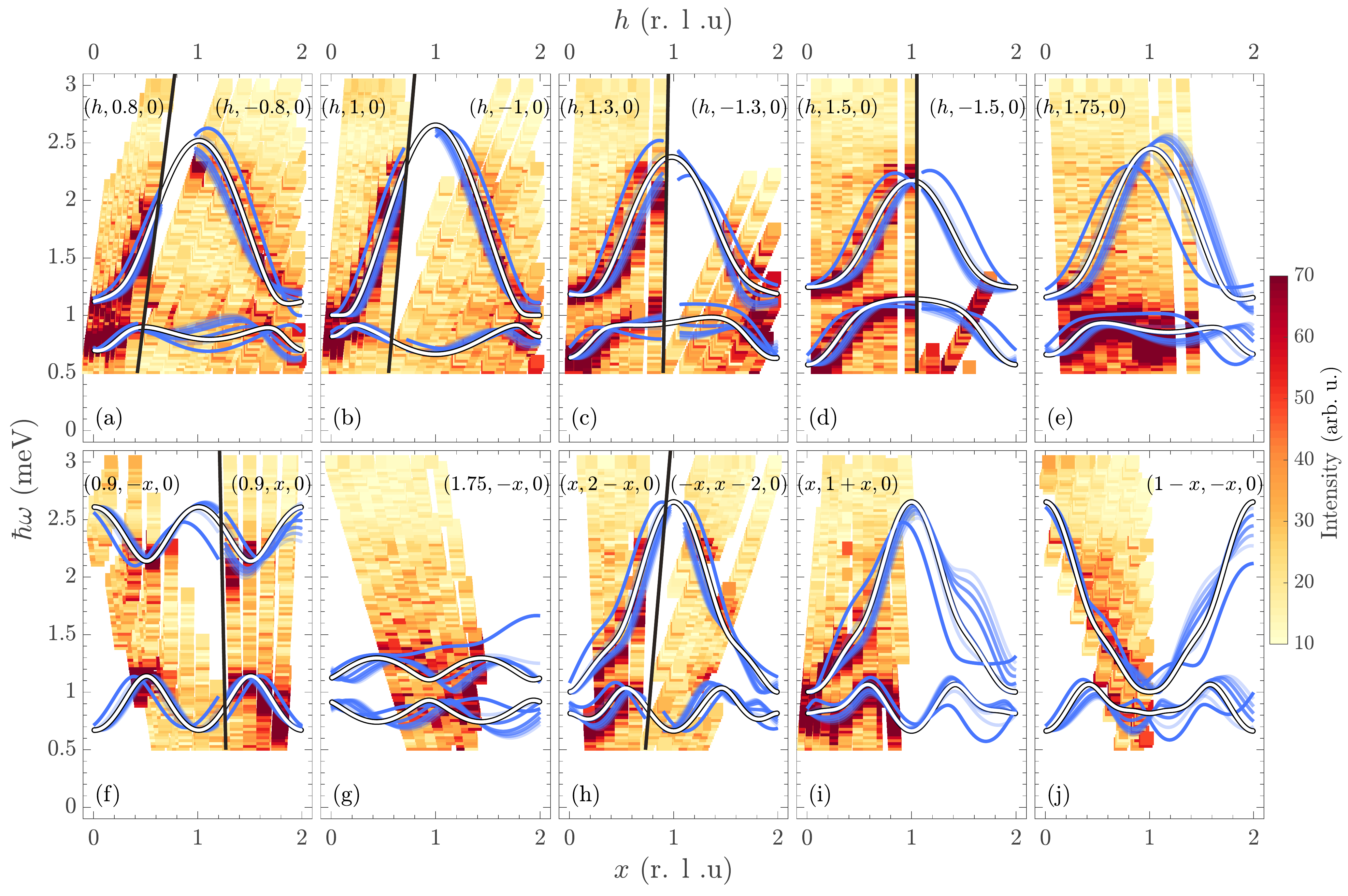}\\
\caption{False color intensity plot of magnetic scattering in \BaCdV shown for several representative slices of energy-momentum space. In all cases the integration range along $h$ or $x$ is $\pm$  0.1~(r. l. u.). The white line is the dispersion relation obtained in a global fit to all collected data, as described in the text. Semi-transparent blue lines are the contributions of individual crystallites. The solid black lines separate data from two different parts of the reciprocal space.}\label{FIG:InelasticScans}
\end{figure*}

 The magnon dispersion relation computed from the obtained exchange constants is represented by white lines Fig.~\ref{FIG:InelasticScans}. Blue lines are contributions of each individual crystallite. In Fig.~\ref{FIG:escans} solid red lines shows results of the global fit and shaded areas are again contributions of individual crystallites.  Considering the global nature of the fit, the complex measured scans profiles are very well reproduced. Any remaining differences may be due to a structured background (e.g., multi-phonon scattering). The possibility that an additional strongly misaligned crystallite was overlooked by our survey can also not be excluded. However, based on the present fit quality we can surmise that the relative weight of the latter must be very small.


\begin{table}
	\centering
	\begin{tabular}{ l r c c c}
		\hline\hline
		Bond     $J$ (meV)       & Length (\AA)\\
		\hline
		$\left. \begin{array}{lr}
			J_{1}^{a}    &-0.135(3) \\         
		    J_{1}^{'a}   &-0.614(4)        
		\end{array}\right\}$&		
		\begin{tabular}{c}
			4.676\\
			4.486        
		\end{tabular}\\

		$\left. \begin{array}{lr}
J_{1}^{b}    &-0.314(6) \\         
J_{1}^{'b}   &-0.464(3)        
\end{array}\right\}$&		
\begin{tabular}{c}
		4.584\\
	4.574        
\end{tabular}\\
\\

		$\left. \begin{array}{lr}
J_{2}^{+}    &0.384(6) \\         
J_{2}^{'+}   &0.039(7)        
\end{array}\right\}$&		
\begin{tabular}{c}
	6.279\\
	6.300        
\end{tabular}\\

$\left. \begin{array}{lr}
J_{2}^{-}    &0.361(5)  \\         
J_{2}^{'-}   &0.181(7)        
\end{array}\right\}$&		
\begin{tabular}{c}
	6.292\\
	6.286        
\end{tabular}\\
\hline\hline
\end{tabular}
	\caption{Parameters of a Heisenberg Hamiltonian for \BaCdV obtained by fitting a spin wave theory model to the entire collected data set. The labeling of exchange parameters are as in Fig.~\ref{FIG:MagneticStr} The corresponding bond distances are also shown.Curly braces indicate that exchange constants can only be assigned to specific crystallographic bonds modulo a petrmutation, as explained in the text.}
\label{tab:exchange}

\end{table}

\section{Discussion and conclusion}
As  expected, \BaCd  is {\it not} the simple $J_1 -J_2$ square lattice material that it was initially believed to be. Instead, it has significantly alternating interactions along the $b$ direction, and also  along the diagonals.  Understanding the structural origin of these variations is challenging. Even for the higher-symmetry room temperature structure  the effective exchange constants represent complex multi-atom superexchange pathways involving distorted oxygen-phosphorous complexes \cite{TsirlinRosner2009}. This said, for nearest-neighbor exchange, we can speculate that a particularly small value of $J_{1}^{a}$ may be associated with the longest $4.676$~\AA\ bond length (see Table~\ref{tab:exchange}). A careful consideration of the 3-dimensional structure reveals that this bond also features the strongest out-of-plane buckling. Structural reasons for a particularly small $J_{2}^{'+}$ are not as obvious.

Despite the variation, NN and NNN interactions are all ferromagnetic and antiferromagnetic, respectively. A quantitative correspondence with the square lattice model can be made by computing the ratio of mean values:
\be
 \frac{\langle J_2 \rangle}{\langle J_1 \rangle} = \frac{J_{2}^{+}+J_{2}^{-}+J_{2}^{'+} + J_{2}^{'-}}{J_{1}^{a}+J_{1}^{'a}+J_{1}^{b}+J_{1}^{'b}} = -0.63.
\ee
The relative strength of ferromagnetic interactions is actually {\em larger} that the $J_2/J_1=-0.9$ estimate from powder studies \cite{Nath2008}, suggesting the system may be more frustrated than originally thought. It is also considerably larger than in the sister compound \PbV where $ \langle J_2 \rangle/\langle J_1 \rangle=- 2.74$ \cite{Bettler2019}.

The minimum of the magnon dispersion computed using the exchange constants listed in Table~\ref{tab:exchange} is located at $\mathbf{q}_\mathrm{min}=(0,1/2,0)$. This exactly corresponds to the propagation vector of the zero-field magnetic structure in \BaCd, which can thus be seen as a magnon condensate. Correspondingly, the computed critical field of single-magnon instability is $\mu_0H_c = 3.92(3) $~T. This is consistent with the experimentally measured field $\mu_0H_c = 4.08(5) $~T, at which the $\mathbf{q} =(0,1/2,0)$ structure collapses \cite{Bhartiya2019}.
We conclude that the previously observed presaturation phase between $\mu_0H_c$ and $\mu_0H_\mathrm{sat} \simeq 6.5$~T is an exotic state from beyond the single-magnon BEC paradigm. As discussed in detail in Refs.~\cite{Smerald2015, Smerald2020}, a spin nematic phase remains a strong candidate. While the present measurement is carried out entirely outside that phase,we hope that the newly obtained model Hamiltonian will help further refine the calculations such as those in Ref.~\cite{Smerald2020}, confirming this expectation.

\acknowledgements

    This work was supported by the Swiss National Science Foundation,
Division II. Access to MACS  was provided by the Center for High Resolution Neutron Scattering, a partnership between the National Institute of Standards and Technology and the National Science Foundation under Agreement No. DMR-1508249. V. B. thanks Florian Landolt for the fruitful discussions.
     \bibliography{The_Library}

\end{document}